\documentclass[a4paper,10pt]{article}

\usepackage{amsmath,amssymb,mathtools}     
\usepackage{color}
\usepackage{graphicx}
\usepackage{subfigure}
\usepackage{cite}                
\usepackage{hyperref}            
\usepackage{multirow,makecell}   
\usepackage{textcomp}
\usepackage{wasysym}

\usepackage[text={17cm,24.5cm},centering]{geometry}

\numberwithin{equation}{section}   

\def \be {\begin{equation}}
\def \ee {\end{equation}}
\def \ba {\begin{array}}
\def \ea {\end{array}}
\def \bea{\begin{eqnarray}}
\def \eea{\end{eqnarray}}
\def \nn {\nonumber}

\def \a {\alpha}
\def \b {\beta}
\def \g {\gamma}

\def \G {\Gamma}

\def \D {\Delta}

\def \r {\rho}

\def \mC {\mathcal C}

\def \mO {\mathcal O}
\def \mP {\mathcal P}

\def \mR {\mathcal R}
\def \mS {\mathcal S}

\def \mU {\mathcal U}
\def \mV {\mathcal V}
\def \mW {\mathcal W}
\def \mX {\mathcal X}
\def \mY {\mathcal Y}
\def \mZ {\mathcal Z}

\def \p {\partial}
\def \f {\frac}

\def \lt {\left}
\def \rt {\right}

\def \inf {\infty}

\def \lag {\langle}
\def \rag {\rangle}

\def \ep {\mathrm{e}}
\def \ii {\mathrm{i}}

\def \and {{\textrm{and}}}

\def \CFT {{\textrm{CFT}}}

\def \vac {{\langle0|}}
\def \uum {{|0\rangle}}

\def \vacu {{\textrm{vac}}}

\def \oloop {{\textrm{1-loop}}}

\begin{document}

\title{\textbf{Note on non-vacuum conformal family contributions to R\'enyi entropy in two-dimensional CFT}}
\author{Jia-ju Zhang\footnote{jiaju.zhang@unimib.it}}
\date{}

\maketitle

\vspace{-10mm}

\begin{center}
{\it
Dipartimento di Fisica, Universit\'a degli Studi di Milano-Bicocca, Piazza della Scienza 3, I-20126 Milano, Italy
}
\vspace{10mm}
\end{center}

\begin{abstract}

  We calculate the contributions of a general non-vacuum conformal family to R\'enyi entropy in two-dimensional conformal field theory (CFT). The primary operator of the conformal family can be either non-chiral or chiral, and we denote its scaling dimension by $\Delta$. For the case of two short intervals on complex plane, we expand the R\'enyi mutual information by the cross ratio $x$ to order $x^{2\Delta+2}$. For the case of one interval on torus with the temperature being low, we expand the R\'enyi entropy by $q=\exp(-2\pi\beta/L)$, with $\beta$ being the inverse temperature and $L$ being the spatial period, to order $q^{\Delta+2}$. To make the result meaningful, we require that the scaling dimension $\Delta$ cannot be too small. For two intervals on complex plane we need $\Delta>1$, and for one interval on torus we need $\Delta>2$. We work in small Newton constant limit in gravity side and so large central charge limit in CFT side, and find matches of gravity and CFT results.

\end{abstract}

\baselineskip 18pt
\thispagestyle{empty}
\newpage

\tableofcontents


\section{Introduction}

The investigation of entanglement entropy has drawn more and more attentions in the past decade, not only because it is interesting in its own right \cite{Calabrese:2004eu}, but also because it opens a new angle in the investigation of AdS/CFT correspondence \cite{Ryu:2006bv,Ryu:2006ef}. To calculate the entanglement entropy in a quantum field theory, one can use the replica trick \cite{Callan:1994py,Holzhey:1994we}. One first calculates the general $n$-th R\'enyi entropy with $n>1$ and being an integer, and then takes the $n \to 1$ limit to get the entanglement entropy. For a conformal field theory (CFT) that has gravity dual in anti-de Sitter (AdS) spacetime \cite{Maldacena:1997re,Gubser:1998bc,Witten:1998qj,Aharony:1999ti}, one can use the Ryu-Takayanagi formula and just calculate the area of a minimal surface in gravity side \cite{Ryu:2006bv,Ryu:2006ef}. The Ryu-Takayanagi area formula of holographic entanglement entropy is the leading classical result in the limit of small Newton constant, and one can also consider the quantum corrections \cite{Fujita:2009kw,Headrick:2010zt,Barrella:2013wja,Faulkner:2013ana}.

In AdS$_3$/CFT$_2$ correspondence, small Newton constant limit in gravity side corresponds to large central charge limit in CFT side \cite{Brown:1986nw}. There are many investigations of R\'enyi entropy and holographic R\'enyi entropy in AdS$_3$/CFT$_2$ correspondence. In gravity side, one uses the partition function of Einstein gravity in handlebody background \cite{Maloney:2007ud,Yin:2007gv,Giombi:2008vd}, and calculates the classical and one-loop parts of the holographic R\'enyi entropy \cite{Faulkner:2013yia,Barrella:2013wja,Chen:2014unl}.
In CFT side one uses different methods to calculate R\'enyi entropy for the cases of two intervals on complex plane and one interval on torus. For the former case, one uses the operator product expansion (OPE) of twist operators \cite{Headrick:2010zt,Calabrese:2010he,Hartman:2013mia,Chen:2013kpa}. For the latter case, one uses the low temperature expansion of density matrix \cite{Cardy:2014jwa,Chen:2014unl}. One can see \cite{Chen:2013dxa,Perlmutter:2013paa,Chen:2014kja,Beccaria:2014lqa,Long:2014oxa,Chen:2014ehg,Chen:2014hta,Chen:2015kua,%
Chen:2015uia,Chen:2015uga,Zhang:2015hoa,Li:2016pwu,Chen:2016uvu,Chen:2016lbu,Li:2016qbo} for other investigations.

In AdS/CFT correspondence different operators in CFT are dual to different fields in gravity side, and it is interesting to compute the contributions of some specific operators to R\'enyi entropy in CFT side and compare the contributions of corresponding fields in gravity side to holographic R\'enyi entropy. The cases of some specific operators have been investigated in literature, for examples stress tensor \cite{Barrella:2013wja,Chen:2013kpa,Chen:2014unl}, $W$ operators \cite{Chen:2013dxa,Perlmutter:2013paa,Chen:2015uia}, logarithmic partner of stress tensor \cite{Chen:2014kja}, general scalars \cite{Beccaria:2014lqa}, supersymmetric partners of stress tensor \cite{Zhang:2015hoa,Li:2016qbo}, and current operator \cite{Li:2016qbo}.
There are also some investigations of the contributions of a general primary operator to R\'enyi entropy \cite{Calabrese:2010he,Barrella:2013wja,Perlmutter:2013paa,Cardy:2014jwa}, and in this paper we generalize the results to higher orders. We consider the contributions of a general non-vacuum conformal family to the R\'enyi entropy, with the primary operator of the conformal family being non-chiral or chiral. The non-chiral primary operator with conformal weights $(h,\bar h)$ has scaling dimension $\D=h+\bar h$, and the chiral primary operator with conformal weights $(h,0)$ has scaling dimension $\D=h$.
For the case of two short intervals on complex plane, we expand the R\'enyi mutual information by the cross ratio $x$ to order $x^{2\Delta+2}$. For the case of one interval on torus with the temperature being low, we expand the R\'enyi entropy by $q=\exp(-2\pi \beta/L)$, with $\b$ being the inverse temperature and $L$ being the spatial period, to order $q^{\D+2}$.

The following part of this paper is arranged as follows. In section~\ref{plane} we consider the R\'enyi mutual information of two intervals on complex plane. In section~\ref{torus} we consider the R\'enyi entropy of one interval on torus. We conclude with discussion in section~\ref{conanddis}. In appendix~\ref{rev} we review some useful properties of the non-vacuum conformal family.

\section{R\'enyi mutual information of two intervals on complex plane}\label{plane}

We calculate the R\'enyi mutual information of two short intervals on complex plane in expansion of the small cross ratio $x$. In gravity side we calculate the one-loop holographic R\'enyi mutual information using the method in \cite{Barrella:2013wja}, and in CFT side we calculate the R\'enyi mutual information using the OPE of twist operators \cite{Headrick:2010zt,Calabrese:2010he,Chen:2013kpa}. In the CFT calculation we will use some results in \cite{Chen:2014kja}.

\subsection{Non-chiral primary operator}

The classical part of the holographic R\'enyi mutual information only depends on the graviton, however the field in gravity dual to a nonidentity primary operator $\mX$ changes the one-loop result.
The non-chiral primary operator $\mX$ has conformal weights $(h,\bar h)$ with $h\neq 0$ and $\bar h\neq 0$, and its conformal weight is $\D=h+\bar h$.
The gravity Euclidean space is the quotient of global AdS$_3$ by a Schottky group $\G$, and the one-loop partition function is multiplied by
\be \label{ee15}
Z_\mX^\oloop =\prod_{\g\in\mP} \bigg(1 + \f{q_\g^h\bar q_\g^{\bar h}}{(1-q_\g)(1-\bar q_\g)}\bigg)^{1/2},
\ee
with $\mP$ being a set of representatives of the primitive conjugacy classes of $\G$. The form of $q_\g$ can be found in \cite{Barrella:2013wja}.
We get the contributions to the one-loop holographic R\'enyi mutual information
\bea \label{InXoloopNonchiral}
&& I_{n,\mX}^\oloop = \f{1}{n-1}\f{x^{2\D}}{2^{4\D+1}n^{4\D-1}}
              \Big\{  f_{2\D}
                     + \f{\D \big( (n^2-1)f_{2\D} + f_{2\D+1} \big)}{n^2}x  \nn\\
&& \phantom{I_{n,\mX}^\oloop =}
                     + \f{1}{72n^4}\big[ \D\big( (36\D+29)(n^2-1)+24 \big)(n^2-1)f_{2\D}
                                         +36\D(2\D+1)(n^2-1)f_{2\D+1} \nn\\
&& \phantom{I_{n,\mX}^\oloop =}
                                         +9(4\D^2+3\D+1)f_{2\D+2} \big] x^2 + O(x^3)
              \Big\} +O(x^{4\D}),
\eea
with the definition
\be
f_m = \sum_{k=1}^{n-1}\f{1}{\big(\sin\f{\pi k}{n}\big)^{2m}}.
\ee
In (\ref{InXoloopNonchiral}) we only incorporated the contributions of the so called consecutively decreasing words (CDW's) of the Schottky generators \cite{Barrella:2013wja}, and the order $x^{4\D}$ result that is omitted is from the 2-CDW's. To make the order $x^{2\D+2}$ part meaningful, we need $4\D>2\D+2$ and so $\D>1$.
Using \cite{Calabrese:2010he}
\be
\lim_{n \to 1}\f{f_m}{n-1}=\f{\sqrt{\pi}\Gamma (m+1)}{2\Gamma (m+{3}/{2})},
\ee
we get the contributions to the one-loop holographic mutual information
\be
I_{\mX}^\oloop =   \frac{\sqrt{\pi} \Gamma(2 \Delta +1) x^{2 \Delta }}
                        {4^{2 \Delta +1}\Gamma(2 \Delta  +3/2)} \Big[ 1+\frac{2 \Delta  (2 \Delta +1) x}{4 \Delta +3}+\frac{(\Delta +1) (2 \Delta +1) ( 4\Delta^2 +3\Delta+1) x^2}{(4 \Delta +3) (4 \Delta +5)}+O(x^3) \Big] +O(x^{4\D}).
\ee
The holographic mutual information is in accord with the result in \cite{Beccaria:2014lqa} when the primary operator $\mX$ is a scaler.

\begin{table}
  \centering
  \begin{tabular}{|c|c|c|c|} \hline
  $L_0+\bar L_0$        & $(L_0,\bar L_0)$ & quasiprimary operators                & degeneracies     \\ \hline

  $\D$                  & $(h,\bar h)$     & $\mX_{j_1}\mX_{j_2}$ with $j_1 < j_2$ & $\f{n(n-1)}{2}$  \\ \hline

  \multirow{2}*{$\D+1$} & $(h+1,\bar h)$   & $\mR_{j_1 j_2}$ with $j_1 < j_2$      & $\f{n(n-1)}{2}$  \\ \cline{2-4}
                        & $(h,\bar h+1)$   & $\mS_{j_1 j_2}$ with $j_1 < j_2$      & $\f{n(n-1)}{2}$  \\ \hline

                        & $(h+2,\bar h)$   & $\mX_{j_1}\mY_{j_2}$ with $j_1 \neq j_2$ & ${n(n-1)}$ \\ \cline{2-4}
                        & $(h,\bar h+2)$   & $\mX_{j_1}\mZ_{j_2}$ with $j_1 \neq j_2$ & ${n(n-1)}$  \\ \cline{2-4}
                        & $(h+1,\bar h+1)$ & $\mW_{j_1j_2}$ with $j_1<j_2$            & $\f{n(n-1)}{2}$  \\ \cline{2-4}
  $\D+2$                & $(h+2,\bar h)$   & $\mU_{j_1j_2}$ with $j_1<j_2$            & $\f{n(n-1)}{2}$  \\ \cline{2-4}
                        & $(h,\bar h+2)$   & $\mV_{j_1j_2}$ with $j_1<j_2$            & $\f{n(n-1)}{2}$  \\ \cline{2-4}
                        & $(h+2,\bar h)$   & $T_{j_1}X_{j_2}X_{j_3}$ with $j_1\neq j_2$, $j_1\neq j_3$ and $j_2<j_3$      & $\f{n(n-1)(n-2)}{2}$ \\ \cline{2-4}
                        & $(h,\bar h+2)$   & $\bar T_{j_1}X_{j_2}X_{j_3}$ with $j_1\neq j_2$, $j_1\neq j_3$ and $j_2<j_3$ & $\f{n(n-1)(n-2)}{2}$ \\ \hline

  $\cdots$              & $\cdots$         & $\cdots$                                 & $\cdots$ \\
\hline
\end{tabular}
  \caption{The quasiprimary operators in $\CFT^n$ from the conformal family of non-chiral primary operator $\mX$ in the original CFT. The indices $j_1$, $j_2$, $j_3$ take values from $0$ to $n-1$.}
  \label{TabNonchiral}
\end{table}

In CFT side, we use the OPE of twist operators in the $n$-fold CFT that is called $\CFT^n$.
The R\'enyi mutual information can be calculated as \cite{Chen:2013kpa,Chen:2013dxa,Perlmutter:2013paa}
\be \label{In}
I_{n} = \f{1}{n-1}\log\Big[ \sum_K \a_K {d_K^2} x^{h_K+\bar h_K} {}_2F_1(h_K,h_K;2h_K;x) {}_2F_1(\bar h_K,\bar h_K;2\bar h_K;x) \Big].
\ee
with $K$ being all the orthogonalized quasiprimary operators $\Phi_K$ in $\CFT^n$.
The coefficients $\a_K$ and $d_K$ are, respectively, the normalization factors and OPE coefficients of $\Phi_K$.
In $\CFT^n$ except the quasiprimary operators that are constructed solely by the vacuum conformal family of the original CFT, we have to consider the extra ones that are listed in Table~\ref{TabNonchiral}. In the table we have the definitions
\bea
&& \mR_{j_1j_2}=\mX_{j_1}\ii\p \mX_{j_2}-\ii\p \mX_{j_1}\mX_{j_2}, ~~
   \mS_{j_1j_2}=\mX_{j_1}\ii\bar\p \mX_{j_2}-\ii\bar\p \mX_{j_1}\mX_{j_2}, \nn\\
&& \mW_{j_1j_2}=\mX_{j_1}\p\bar\p \mX_{j_2}+\p\bar\p \mX_{j_1} \mX_{j_2}-\p \mX_{j_1}\bar\p \mX_{j_2}-\bar\p \mX_{j_1}\p \mX_{j_2}, \nn\\
&& \mU_{j_1j_2}=\p \mX_{j_1}\p \mX_{j_2}-\f{h}{2h+1} \lt( \mX_{j_1}\p^2 \mX_{j_2} + \p^2 \mX_{j_1} \mX_{j_2} \rt),\\
&& \mV_{j_1j_2}=\bar\p \mX_{j_1}\bar\p \mX_{j_2}-\f{\bar h}{2\bar h+1} \lt( \mX_{j_1}\bar\p^2 \mX_{j_2} + \bar\p^2 \mX_{j_1} \mX_{j_2} \rt).\nn
\eea
We get contributions to R\'enyi mutual information from conformal family of $\mX$
\bea \label{InXNonchiral}
&& I_{n,\mX} = \f{x^{2\D}}{n-1} \Big\{
    \big[ 1-n x^2( \a_T d_T^2 + \a_{\bar T} d_{\bar T}^2) \big] {}_2F_1(2h,2h;4h;x) {}_2F_1(2\bar h,2\bar h;4\bar h;x) \sum \a_{\mX\mX} \big(d_{\mX\mX}^{j_1j_2}\big)^2   \nn\\
&& \phantom{I_{n,\mX} =}
+ x \Big[ {}_2F_1(2h+1,2h+1;4h+2;x) {}_2F_1(2\bar h,2\bar h;4\bar h;x)\sum \a_\mR \big(d_\mR^{j_1j_2}\big)^2   \nn\\
&& \phantom{I_{n,\mX} = + x \big[}
          {}_2F_1(2h,2h;4h;x) {}_2F_1(2\bar h+1,2\bar h+1;4\bar h+2;x)\sum \a_\mS \big(d_\mS^{j_1j_2}\big)^2
    \Big] \\
&& \phantom{I_{n,\mX} =}
+x^2 \Big[  \sum \Big(\a_{\mX\mY} \big(d_{\mX\mY}^{j_1j_2}\big)^2+\a_{\mX\mZ} \big(d_{\mX\mZ}^{j_1j_2}\big)^2\Big)
          + \sum \Big( \a_\mW \big(d_\mW^{j_1j_2}\big)^2
                     + \a_\mU \big(d_\mU^{j_1j_2}\big)^2
                     + \a_\mV \big(d_\mV^{j_1j_2}\big)^2 \Big)   \nn\\
&& \phantom{I_{n,\mX} =+x^2}
         + \sum \Big(\a_{T\mX\mX} \big(d_{T\mX\mX}^{j_1j_2j_3}\big)^2+\a_{\bar T\mX\mX} \big(d_{\bar T\mX\mX}^{j_1j_2j_3}\big)^2\Big) \Big] + O(x^3)
\Big\} + O(x^{3\D}), \nn
\eea
and the ranges of summations can be found in Table~\ref{TabNonchiral}. The order $x^{3\D}$ result that is omitted in the above result is from contributions of the $\CFT^n$ operators $\mX_{j_1}\mX_{j_2}\mX_{j_3}$ with $0\leq j_1<j_2<j_3\leq n-1$.

We have the normalization factors \cite{Chen:2014kja}
\bea
&& \a_{\mX\mX}=\ii^{4s}\a_\mX^2, ~~
   \a_\mR=4h \ii^{4s}\a_\mX^2, ~~
   \a_\mS=4\bar h  \ii^{4s}\a_\mX^2, ~~
   \a_{\mX \mY}=\f{(2h+1)c+2h(8h-5)}{2(2h+1)} \ii^{4s}\a_\mX^2, \nn\\
&& \a_{\mX\mZ}=\f{(2\bar h+1)c+2\bar h(8\bar h-5)}{2(2\bar h+1)} \ii^{4s}\a_\mX^2, ~~
   \a_\mW=16h\bar h \ii^{4s}\a_\mX^2, ~~
   \a_\mU=\f{4h^2(4h+1)}{2h+1} \ii^{4s}\a_\mX^2,\\
&& \a_\mV=\f{4\bar h^2(4\bar h+1)}{2\bar h+1} \ii^{4s}\a_\mX^2, ~~
   \a_{T\mX\mX}=\f{c}{2} \ii^{4s}\a_\mX^2, ~~
   \a_{\bar T\mX\mX}=\f{c}{2} \ii^{4s}\a_\mX^2,\nn
\eea
where the factor $\ii^{4s}=(-1)^{2s}$ aries from the minus sign when $\mX$ is an fermionic operator. Note that there is always $\ii^{8s}=1$. We also have the OPE coefficients \cite{Chen:2014kja}
\bea
&& d_{\mX\mX}^{j_1j_2}=\f{\ii^{2s}}{\a_\mX(2n)^{2\D}}\f{1}{s_{j_1j_2}^{2\D}}, ~~
   d_\mR^{j_1j_2}=-d_\mS^{j_1j_2}=\f{\ii^{2s}}{\a_\mX(2n)^{2\D+1}}\f{c_{j_1j_2}}{s_{j_1j_2}^{2\D+1}},   \nn\\
&& d_{\mX\mY}^{j_1j_2}=d_{\mX\mZ}^{j_1j_2}=\f{\ii^{2s}(n^2-1)}{3\a_\mX(2n)^{2\D+2}}\f{1}{s_{j_1j_2}^{2\D}},  ~~
   d_\mW^{j_1j_2}=\f{\ii^{2s}}{\a_\mX(2n)^{2\D+2}} \f{c_{j_1j_2}^2}{s_{j_1j_2}^{2\D+2}},   \nn\\
&& d_\mU^{j_1j_2}=\f{\ii^{2s}}{2h(4h+1)\a_\mX(2n)^{2\D+2}}\f{(2h+1)(4h+1)-2h(n^2+4h+1)s_{j_1j_2}^2}{s_{j_1j_2}^{2\D+2}},  \nn\\
&& d_\mV^{j_1j_2}=\f{\ii^{2s}}{2\bar h(4\bar h+1)\a_\mX(2n)^{2\D+2}}\f{(2\bar h+1)(4\bar h+1)-2\bar h(n^2+4\bar h+1)s_{j_1j_2}^2}{s_{j_1j_2}^{2\D+2}},\\
&& d_{T\mX\mX}^{j_1j_2j_3}=\f{\ii^{2s}}{\a_\mX(2n)^{2\D+2}} \Big( -\f{2h}{c} \f{1}{s_{j_1j_2}^2s_{j_1j_3}^2s_{j_2j_3}^{2\D-2}}
                                                         +\f{n^2-1}{3}\f{1}{s_{j_2j_3}^{2\D}} \Big),\nn\\
&& d_{\bar T\mX\mX}^{j_1j_2j_3}=\f{\ii^{2s}}{\a_\mX(2n)^{2\D+2}} \Big( -\f{2\bar h}{c} \f{1}{s_{j_1j_2}^2s_{j_1j_3}^2s_{j_2j_3}^{2\D-2}}
                                                         +\f{n^2-1}{3}\f{1}{s_{j_2j_3}^{2\D}} \Big). \nn
\eea
Here for simplicity we have defined $s_{j_1j_2}\equiv\sin\f{\pi(j_1-j_2)}{n}$, $c_{j_1j_2}\equiv\cos\f{\pi(j_1-j_2)}{n}$, $\cdots$. Besides, we also need the normalization factors and OPE coefficients for the operators $T_j$ and $\bar T_j$ with $j=0,1,\cdots,n-1$
\be
\a_T=\a_{\bar T}=\f{c}{2}, ~~
d_T = d_{\bar T} = \frac{n^2-1}{12n^2}.
\ee
With these coefficients and the formula (\ref{InXNonchiral}), in large $c$ limit we can reproduce the one-loop holomorphic R\'enyi mutual information (\ref{InXoloopNonchiral}).

\subsection{Chiral primary operator}

The case of chiral primary operator $\mX$, with conformal weights $(h,0)$ and $h \neq 0$, is similar to but a little different from the non-chiral operator case, and we discuss this briefly in the subsection. Note that we only consider the contributions of the conformal family $\mX$, and we do not count the contributions of the possible conformal family of the anti-holomorphic operator $\bar \mX$ with conformal weights $(0,h)$.

Similar to (\ref{ee15}), the one-loop partition function is multiplied by
\be \label{ee16}
Z_\mX^\oloop =\prod_{\g\in\mP} \bigg(1 + \f{q_\g^h}{1-q_\g}\bigg)^{1/2}.
\ee
We get the contributions to the one-loop holographic R\'enyi mutual information
\bea \label{InXoloopChiral}
&& I_{n,\mX}^\oloop = \f{1}{n-1}\f{x^{2h}}{2^{4h+1}n^{4h-1}}
              \Big\{  f_{2h}
                     + \f{h \big( (n^2-1)f_{2h} + f_{2h+1} \big)}{n^2}x  \nn\\
&& \phantom{I_{n,\mX}^\oloop =}
                     + \f{1}{144n^4}\big[ 2h\big( (36h+29)(n^2-1)+24 \big)(n^2-1)f_{2h}
                                         +72h(2h+1)(n^2-1)f_{2h+1} \nn\\
&& \phantom{I_{n,\mX}^\oloop =}
                                         +9(8h^2+6h+1)f_{2h+2} \big] x^2 + O(x^3)
              \Big\} +O(x^{4h}),
\eea
as well as the one-loop holographic mutual information
\be
I_{\mX}^\oloop =   \frac{\sqrt{\pi} \Gamma(2 h +1) x^{2 h }}{4^{2 h +1}\Gamma(2 h  +3/2)}
       \Big[ 1+\frac{2 h (2 h+1) x}{4 h+3}+\frac{(h+1) (2 h+1)^2 (4 h+1) x^2}{2 (4 h+3) (4 h+5)}+O(x^3) \Big] +O(x^{4h}).
\ee
The holographic mutual information is in accord with the result in \cite{Beccaria:2014lqa}.

\begin{table}
  \centering
  \begin{tabular}{|c|c|c|} \hline
  $L_0$ & quasiprimary operators                   & degeneracies     \\ \hline

  $h$   & $\mX_{j_1}\mX_{j_2}$ with $j_1 < j_2$    & $\f{n(n-1)}{2}$  \\ \hline

  $h+1$ & $\mR_{j_1 j_2}$ with $j_1 < j_2$         & $\f{n(n-1)}{2}$  \\ \hline

        & $\mX_{j_1}\mY_{j_2}$ with $j_1 \neq j_2$ & $n(n-1)$ \\ \cline{2-3}
  $h+2$ & $\mU_{j_1j_2}$ with $j_1<j_2$            & $\f{n(n-1)}{2}$  \\ \cline{2-3}
        & $T_{j_1}X_{j_2}X_{j_3}$ with $j_1\neq j_2$, $j_1\neq j_3$ and $j_2<j_3$ & $\f{n(n-1)(n-2)}{2}$ \\ \hline

  $\cdots$ & $\cdots$ & $\cdots$ \\
\hline
\end{tabular}
  \caption{The quasiprimary operators in $\CFT^n$ from the conformal family of chiral primary operator $\mX$ in the original CFT.}
  \label{TabChiral}
\end{table}

In CFT side, we have to consider the extra quasiprimary operators that are listed in Table~\ref{TabChiral}. In the table we have definitions
\be
\mR_{j_1j_2}=\mX_{j_1}\ii\p \mX_{j_2}-\ii\p \mX_{j_1}\mX_{j_2}, ~~
\mU_{j_1j_2}=\p \mX_{j_1}\p \mX_{j_2}-\f{h}{2h+1} \lt( \mX_{j_1}\p^2 \mX_{j_2} + \p^2 \mX_{j_1} \mX_{j_2} \rt).
\ee
We get contributions to R\'enyi mutual information from conformal family $\mX$
\bea \label{InXChiral}
&& I_{n,\mX} = \f{x^{2h}}{n-1} \Big\{
    \big[ 1 - x^2 n \a_T d_T^2 \big] {}_2F_1(2h,2h;4h;x) \sum \a_{\mX\mX} \big(d_{\mX\mX}^{j_1j_2}\big)^2 \nn\\
&& \phantom{I_{n,\mX} =}
         + x {}_2F_1(2h+1,2h+1;4h+2;x) \sum \a_\mR \big(d_\mR^{j_1j_2}\big)^2   \\
&& \phantom{I_{n,\mX} =}
+x^2 \Big[  \sum \a_{\mX\mY} \big(d_{\mX\mY}^{j_1j_2}\big)^2
          + \sum \a_\mU \big(d_\mU^{j_1j_2}\big)^2
          + \sum \a_{T\mX\mX} \big(d_{T\mX\mX}^{j_1j_2j_3}\big)^2 \Big] + O(x^3)
\Big\} + O(x^{3h}). \nn
\eea
We have the normalization factors
\bea
&& \a_{\mX\mX}=\ii^{4h}\a_\mX^2, ~~
\a_\mR=4h \ii^{4h}\a_\mX^2, ~~
\a_{\mX \mY}=\f{(2h+1)c+2h(8h-5)}{2(2h+1)} \ii^{4h}\a_\mX^2, \nn\\
&& \a_\mU=\f{4h^2(4h+1)}{2h+1} \ii^{4h}\a_\mX^2, ~~
\a_{T\mX\mX}=\f{c}{2} \ii^{4h}\a_\mX^2,
\eea
and  the OPE coefficients
\bea
&& d_{\mX\mX}^{j_1j_2}=\f{\ii^{2h}}{\a_\mX(2n)^{2h}}\f{1}{s_{j_1j_2}^{2h}}, ~~
   d_\mR^{j_1j_2}=\f{\ii^{2h}}{\a_\mX(2n)^{2h+1}}\f{c_{j_1j_2}}{s_{j_1j_2}^{2h+1}}, ~~
   d_{\mX\mY}^{j_1j_2}=\f{\ii^{2h}(n^2-1)}{3\a_\mX(2n)^{2h+2}}\f{1}{s_{j_1j_2}^{2h}}, \nn\\
&& d_\mU^{j_1j_2}=\f{\ii^{2h}}{2h(4h+1)\a_\mX(2n)^{2h+2}}\f{(2h+1)(4h+1)-2h(n^2+4h+1)s_{j_1j_2}^2}{s_{j_1j_2}^{2h+2}},  \\
&& d_{T\mX\mX}^{j_1j_2j_3}=\f{\ii^{2h}}{\a_\mX(2n)^{2h+2}} \Big( -\f{2h}{c} \f{1}{s_{j_1j_2}^2s_{j_1j_3}^2s_{j_2j_3}^{2h-2}}
                                                         +\f{n^2-1}{3}\f{1}{s_{j_2j_3}^{2h}} \Big).\nn
\eea
Using the formula (\ref{InXChiral}), we can reproduce the one-loop holomorphic R\'enyi mutual information (\ref{InXoloopChiral}).

\section{R\'enyi entropy of one interval on torus}\label{torus}

We calculate the contributions of a non-vacuum conformal family to R\'enyi entropy of one interval with length $\ell$ on torus in low temperature limit. The torus has spatial period $L$ and temporal period $\b$, with the temperature being $1/\b$, and in low temperature we have $\b/L \gg 1$.
In gravity side we use the method in \cite{Barrella:2013wja,Chen:2014unl}, and in CFT side we use the method in \cite{Cardy:2014jwa,Chen:2014unl}.

\subsection{Non-chiral primary operator}

In gravity side the one-loop partition function (\ref{ee15}) still applies, and we use a different Schottky group that can be found in \cite{Barrella:2013wja,Chen:2014unl}.
We get the contributions of a non-chiral conformal family to the one-loop holographic R\'enyi entropy
\bea \label{SnXgravity}
&& S_{n,\mX}^\oloop= -\f{n q^\D}{n-1} \bigg\{ \bigg[ \f{1}{n^{2\D}}\bigg(\f{\sin\f{\pi\ell}{L}}{\sin\f{\pi\ell}{nL}}\bigg)^{2\D} -1\bigg]  \nn\\
&& \phantom{S_{n,\mX}^\oloop=}
+\bigg[ \f{2}{n^{2\D+2}} \bigg( \f{\sin\f{\pi\ell}{L}}{\sin\f{\pi\ell}{nL}}\bigg)^{2\D} \bigg(
       n^2\D\cos^2\f{\pi\ell}{L}
      -n\D\sin\f{2\pi\ell}{L}\cot\f{\pi\ell}{nL}
      +\f{\sin^2\f{\pi\ell}{L}}{\sin^2\f{\pi\ell}{nL}} \Big(\D\cos^2\f{\pi\ell}{nL}+1\Big)  \bigg)-2\bigg]q                       \nn\\
&& \phantom{S_{n,\mX}^\oloop=}
+\bigg[ \f{1}{9n^{2\D+4}}\bigg(\f{\sin\f{\pi\ell}{L}}{\sin\f{\pi\ell}{nL}}\bigg)^{2\D}\bigg(
        n^4\D \Big( (18\D+29)\cos^4\f{\pi\ell}{L}-16\cos^2\f{\pi\ell}{L}-4\Big)                                                   \nn\\
&& \phantom{S_{n,\mX}^\oloop=}
      -6n^3\D \sin\f{2\pi\ell}{L}\cot\f{\pi\ell}{nL} \Big( (6\D+5)\cos^2\f{\pi\ell}{L}-2\Big)                                     \\
&& \phantom{S_{n,\mX}^\oloop=}
      +2n^2 \f{\sin^2\f{\pi\ell}{L}}{\sin^2\f{\pi\ell}{nL}}\Big( \D(54\D+19)\cos^2\f{\pi\ell}{L}\cos^2\f{\pi\ell}{nL}
                                                                +(35\D+18)\cos^2\f{\pi\ell}{L}
                                                                +4\D\sin^2\f{\pi\ell}{nL} \Big)                                   \nn\\
&& \phantom{S_{n,\mX}^\oloop=}
     -12n \f{\sin^3\f{\pi\ell}{L}}{\sin^3\f{\pi\ell}{nL}}\cos\f{\pi\ell}{L}\cos\f{\pi\ell}{nL} \Big( \D(6\D+1)\cos^2\f{\pi\ell}{nL}+11\D+6\Big) \nn\\
&& \phantom{S_{n,\mX}^\oloop=}
    +\f{\sin^4\f{\pi\ell}{L}}{\sin^4\f{\pi\ell}{nL}}\Big( \D(18\D+5)\cos^4\f{\pi\ell}{nL}+2(31\D+18)\cos^2\f{\pi\ell}{nL}-4\D+27\Big)
\bigg)-3\bigg]q^2 +O(q^3) \bigg\} + O(q^{2\D}),  \nn
\eea
with $q=\ep^{-2\pi\b/L} \ll 1$.
In (\ref{SnXgravity}) the omitted order $q^{2\D}$ result is from the 2-CDW's \cite{Barrella:2013wja}.
To make the order $q^{\D+2}$ part meaningful, we need $2\D>\D+2$ and so $\D>2$.
Taking $n \to 1$ limit, we get the one-loop holographic entanglement entropy
\be
S_{\mX}^\oloop = \Big( 1-\f{\pi\ell}{L}\cot\f{\pi\ell}{L} \Big) 2 q^\D \big( \D +2(\D+1)q +3(\D+2)q^2+O(q^3) \big) + O(q^{2\D}).
\ee

Then we calculate the R\'enyi entropy in CFT side. For the vacuum conformal family we have the density matrix
\be
\r_\vacu = \uum \vac +  \f{q^2}{\a_T}|T\rag\lag T| + \f{q^2}{\a_{\bar T}}|\bar T\rag\lag \bar T|  + O(q^3).
\ee
Note that we only consider the case without chemical potential. Considering the contributions of the conformal family of a non-chiral primary operator $\mX$ we have the density matrix
\be
\r = \r_\vacu + \r_\mX,
\ee
with
\bea
&& \r_\mX = q^\D \Big(  \f{1}{\a_\mX}|\mX \rag \lag \mX |
                   + \f{q}{\a_{\p\mX}}|\p\mX \rag \lag \p\mX |
                   + \f{q}{\a_{\bar\p\mX}}|\bar\p\mX \rag \lag\bar\p\mX |
                   + \f{q^2}{\a_\mY}|\mY \rag \lag \mY |
                   + \f{q^2}{\a_\mZ}|\mZ \rag \lag \mZ | \nn\\
&& \phantom{\r_\mX =}
                   + \f{q^2}{\a_{\p \bar \p \mX}}|\p \bar \p \mX \rag \lag \p \bar \p \mX |
                   + \f{q^2}{\a_{\p^2\mX}}|\p^2\mX \rag \lag \p^2\mX |
                   + \f{q^2}{\a_{\bar\p^2\mX}}|\bar\p^2\mX \rag \lag \bar\p^2\mX | + O(q^3)  \Big).
\eea
In CFT side, we get the R\'enyi entropy
\bea
&& S_{n,\mX}= -\f{n q^\D}{n-1} \Big\{ \Big( \f{\lag\mX(\inf,\inf)\mX(0,0)\rag_{\mC^n}}{\a_\mX} -1 \Big)
   +\Big( \f{\lag\p\mX(\inf,\inf)\p\mX(0,0)\rag_{\mC^n}}{\a_{\p\mX}}
          +\f{\lag\bar\p\mX(\inf,\inf)\bar\p\mX(0,0)\rag_{\mC^n}}{\a_{\bar\p\mX}} -2\Big)q                       \nn\\
&& \phantom{S_{n,\mX}=}
   +\Big[ \f{\lag\p^2\mX(\inf,\inf)\p^2\mX(0,0)\rag_{\mC^n}}{\a_{\p^2\mX}}
          +\f{\lag\p\bar\p\mX(\inf,\inf)\p\bar\p\mX(0,0)\rag_{\mC^n}}{\a_{\p\bar\p\mX}}
          +\f{\lag\bar\p^2\mX(\inf,\inf)\bar\p^2\mX(0,0)\rag_{\mC^n}}{\a_{\bar\p^2\mX}} \nn\\
&& \phantom{S_{n,\mX}=}
          +\f{\lag\mY(\inf,\inf)\mY(0,0)\rag_{\mC^n}}{\a_{\mY}}
          -n\f{\lag T(\inf)T(0)\rag_{\mC^n}\lag\mX(\inf,\inf)\mX(0,0)\rag_{\mC^n}}{\a_{T}\a_{\mX}} \nn\\
&& \phantom{S_{n,\mX}=}
          +\f{\lag\mZ(\inf,\inf)\mZ(0,0)\rag_{\mC^n}}{\a_{\mZ}}
          -n\f{\lag \bar T(\inf)\bar T(0)\rag_{\mC^n}\lag\mX(\inf,\inf)\mX(0,0)\rag_{\mC^n}}{\a_{\bar T}\a_{\mX}} \\
&& \phantom{S_{n,\mX}=}
          +\sum_{j=1}^{n-1} \Big(  \f{\lag T(\inf)T(0)\mX(\inf_j,\inf_j)\mX(0_j,0_j)\rag_{\mC^n}}{\a_{T}\a_{\mX}}
                                   +\f{\lag \bar T(\inf)\bar T(0)\mX(\inf_j,\inf_j)\mX(0_j,0_j)\rag_{\mC^n}}{\a_{\bar T}\a_{\mX}} \Big)
          -3\Big]q^2 \nn\\
&& \phantom{S_{n,\mX}=}
   +O(q^3) \Big\} + O(q^{2\D}).                                                                                   \nn
\eea
Note that in $\inf_j$ and $0_j$ the subscript $j$ denotes different replicas, and the $\inf$ and $0$ without any subscript mean the special $j=0$ case.
The correlation functions on the $n$-fold complex plane $\mC^n$ can be calculated by mapping it to an ordinary complex plane $\mC$ by a conformal transformation. The results are
\bea
&& \f{\lag\mX(\inf,\inf)\mX(0,0)\rag_{\mC^n}}{\a_\mX}=\f{1}{n^{2\D}}\bigg(\f{\sin\f{\pi\ell}{L}}{\sin\f{\pi\ell}{nL}}\bigg)^{2\D},  \nn\\
&& \f{\lag\p\mX(\inf,\inf)\p\mX(0,0)\rag_{\mC^n}}{\a_{\p\mX}}=
     \f{2}{n^{2\D+2}} \bigg( \f{\sin\f{\pi\ell}{L}}{\sin\f{\pi\ell}{nL}}\bigg)^{2\D}
     \bigg( n^2h\cos^2\f{\pi\ell}{L}
           -nh\sin\f{2\pi\ell}{L}\cot\f{\pi\ell}{nL}
           +\f{\sin^2\f{\pi\ell}{L}}{\sin^2\f{\pi\ell}{nL}} \Big(h\cos^2\f{\pi\ell}{nL}+\f12\Big)  \bigg),                                \nn\\
&& \f{\lag\bar\p\mX(\inf,\inf)\bar\p\mX(0,0)\rag_{\mC^n}}{\a_{\bar\p\mX}}
               =\f{\lag\p\mX(\inf,\inf)\p\mX(0,0)\rag_{\mC^n}}{\a_{\p\mX}} \Big|_{h\to\bar h},                                            \nn\\
&& \f{\lag\p^2\mX(\inf,\inf)\p^2\mX(0,0)\rag_{\mC^n}}{\a_{\p^2\mX}}
      =\f{1}{2h+1}\f{1}{n^{2\D+4}}\bigg( \f{\sin\f{\pi\ell}{L}}{\sin\f{\pi\ell}{nL}}\bigg)^{2\D} \bigg[
          n^4h \Big( 2(h+1)\cos^2\f{\pi\ell}{L}-1\Big)^2                                                                                  \nn\\
&& \phantom{\f{\lag\p^2\mX(\inf,\inf)\p^2\mX(0,0)\rag_{\mC^n}}{\a_{\p^2\mX}}=}
         -2n^3h(2h+1) \sin\f{2\pi\ell}{L}\cot\f{\pi\ell}{nL} \Big( 2(h+1)\cos^2\f{\pi\ell}{L}-1\Big)                                      \nn\\
&& \phantom{\f{\lag\p^2\mX(\inf,\inf)\p^2\mX(0,0)\rag_{\mC^n}}{\a_{\p^2\mX}}=}
         +2n^2 \f{\sin^2\f{\pi\ell}{L}}{\sin^2\f{\pi\ell}{nL}}\Big( 2h\cos^2\f{\pi\ell}{nL}+1\Big)
                                                              \Big( (6h^2+6h+1)\cos^2\f{\pi\ell}{L}-h\Big)                                \nn\\
&& \phantom{\f{\lag\p^2\mX(\inf,\inf)\p^2\mX(0,0)\rag_{\mC^n}}{\a_{\p^2\mX}}=}
        -4n(2h+1)\f{\sin^3\f{\pi\ell}{L}}{\sin^3\f{\pi\ell}{nL}}\cos\f{\pi\ell}{L}\cos\f{\pi\ell}{nL}
                 \Big( 2h^2\cos^2\f{\pi\ell}{nL}+3h+1\Big) \nn\\
&& \phantom{\f{\lag\p^2\mX(\inf,\inf)\p^2\mX(0,0)\rag_{\mC^n}}{\a_{\p^2\mX}}=}
        +\f{\sin^4\f{\pi\ell}{L}}{\sin^4\f{\pi\ell}{nL}}\Big( 4h^3\cos^4\f{\pi\ell}{nL}+2(6h^2+4h+1)\cos^2\f{\pi\ell}{nL}+3h+1\Big)\bigg], \nn\\
&& \f{\lag\bar\p^2\mX(\inf,\inf)\bar\p^2\mX(0,0)\rag_{\mC^n}}{\a_{\bar\p^2\mX}}
            =\f{\lag\p^2\mX(\inf,\inf)\p^2\mX(0,0)\rag_{\mC^n}}{\a_{\p^2\mX}} \Big|_{h\to\bar h}, 
\eea\bea
&& \f{\lag\p\bar\p\mX(\inf,\inf)\p\bar\p\mX(0,0)\rag_{\mC^n}}{\a_{\p\bar\p\mX}}
       =\f{\lag\p\mX(\inf,\inf)\p\mX(0,0)\rag_{\mC^n}}{\a_{\p\mX}}
        \f{\lag\bar\p\mX(\inf,\inf)\bar\p\mX(0,0)\rag_{\mC^n}}{\a_{\bar\p\mX}}
        \f{\a_{\mX}}{\lag\mX(\inf,\inf)\mX(0,0)\rag_{\mC^n}},                                                                              \nn\\
&& \f{\lag\mY(\inf,\inf)\mY(0,0)\rag_{\mC^n}}{\a_{\mY}} = \f{1}{n^{2\D+4}}\bigg( \f{\sin\f{\pi\ell}{L}}{\sin\f{\pi\ell}{nL}}\bigg)^{2\D}
        \bigg[ \f{(n^2-1)^2}{18(2h+1)}\sin^4\f{\pi\ell}{L}
               \Big( (2h+1)c+2h(8h-5) \Big) + \f{\sin^4\f{\pi\ell}{L}}{\sin^4\f{\pi\ell}{nL}} \bigg],                                      \nn\\
&& \f{\lag\mZ(\inf,\inf)\mZ(0,0)\rag_{\mC^n}}{\a_{\mZ}}=\f{\lag\mY(\inf,\inf)\mY(0,0)\rag_{\mC^n}}{\a_{\mY}}\Big|_{h\to\bar h}             \nn\\
&& \f{\lag T(\inf)T(0)\mX(\inf_j,\inf_j)\mX(0_j,0_j)\rag_{\mC^n}}{\a_{T}\a_{\mX}}
      = \f{1}{n^{2\D+4}}\bigg( \f{\sin\f{\pi\ell}{L}}{\sin\f{\pi\ell}{nL}}\bigg)^{2\D} \sin^4\f{\pi\ell}{L}
        \bigg[ \f{(n^2-1)^2c}{18} + \f{1}{\sin^4\f{\pi\ell}{nL}}  \nn\\
&& \phantom{\f{ \lag T(\inf)T(0)\mX(\inf_j,\inf_j)\mX(0_j,0_j)\rag_{\mC^n}}{\a_{T}\a_{\mX}}=}
               +\f{(n^2-1)h}{3}\sin^2\f{\pi\ell}{nL}
                \f{\cos\f{2\pi j}{n}\cos\f{2\pi\ell}{nL}-1}{\sin^2\f{\pi j}{n}\lt(\cos\f{2\pi j}{n}-\cos\f{2\pi\ell}{nL}\rt)^2} \bigg] + O\Big(\f{1}{c}\Big),   \nn\\
&& \f{\lag\bar T(\inf)\bar T(0)\mX(\inf_j,\inf_j)\mX(0_j,0_j)\rag_{\mC^n}}{\a_{\bar T}\a_{\mX}}
     = \f{\lag T(\inf)T(0)\mX(\inf_j,\inf_j)\mX(0_j,0_j)\rag_{\mC^n}}{\a_{T}\a_{\mX}}\Big|_{h\to\bar h}.                                  \nn
\eea
In the last two correlation functions we have omitted the $O(1/c)$ terms in large $c$ limit.
Also we have to use \cite{Chen:2014unl}
\be
\f{\lag T(\inf)T(0)\rag_{\mC^n}}{\a_T} =
\f{\lag \bar T(\inf)\bar T(0)\rag_{\mC^n}}{\a_{\bar T}} =
\f{c(n^2-1)^2}{18n^4}\sin^4 \f{\pi\ell}{L} + \f{1}{n^{4}}\f{\sin^4\f{\pi\ell}{L}}{\sin^4\f{\pi\ell}{nL}}.
\ee
Given the summation formula
\be
\sum_{j=1}^{n-1} \f{\cos\f{2\pi j}{n}\cos\f{2\pi\ell}{nL}-1}{\sin^2\f{\pi j}{n}\lt(\cos\f{2\pi j}{n}-\cos\f{2\pi\ell}{nL}\rt)^2}
   =  \f{n^2\lt(\cos^2\f{\pi\ell}{L}-4\rt)}{6\sin^2\f{\pi\ell}{L}\sin^2\f{\pi\ell}{nL}}
     -\f{n\cos\f{\pi\ell}{L}\cos\f{\pi\ell}{nL}}{\sin\f{\pi\ell}{L}\sin^3\f{\pi\ell}{nL}}
     +\f{5\cos^2\f{\pi\ell}{nL}+4}{6\sin^4\f{\pi\ell}{nL}},
\ee
we reproduce the one-loop gravity result (\ref{SnXgravity}) in large $c$ limit.

\subsection{Chiral primary operator}

Using the one-loop partition function (\ref{ee16}) and the Schottky group in \cite{Barrella:2013wja,Chen:2014unl}, we get the contributions of the fields dual to the conformal family of a chiral primary operator to the one-loop holographic R\'enyi entropy
\bea \label{SnXgravityChiral}
&& S_{n,\mX}^\oloop= -\f{n q^h}{n-1} \bigg\{ \bigg[ \f{1}{n^{2h}}\bigg(\f{\sin\f{\pi\ell}{L}}{\sin\f{\pi\ell}{nL}}\bigg)^{2h} -1\bigg]  \nn\\
&& \phantom{S_{n,\mX}^\oloop=}
+\bigg[ \f{2}{n^{2h+2}} \bigg( \f{\sin\f{\pi\ell}{L}}{\sin\f{\pi\ell}{nL}}\bigg)^{2h} \bigg(
       n^2h\cos^2\f{\pi\ell}{L}
      -nh\sin\f{2\pi\ell}{L}\cot\f{\pi\ell}{nL}
      +\f{\sin^2\f{\pi\ell}{L}}{\sin^2\f{\pi\ell}{nL}} \Big(h\cos^2\f{\pi\ell}{nL}+\f12 \Big)  \bigg)-1\bigg]q                       \nn\\
&& \phantom{S_{n,\mX}^\oloop=}
+\bigg[ \f{1}{9n^{2h+4}}\bigg(\f{\sin\f{\pi\ell}{L}}{\sin\f{\pi\ell}{nL}}\bigg)^{2h}\bigg(
        n^4h \Big( (18h+29)\cos^4\f{\pi\ell}{L}-16\cos^2\f{\pi\ell}{L}-4\Big)                                                   \nn\\
&& \phantom{S_{n,\mX}^\oloop=}
      -6n^3h \sin\f{2\pi\ell}{L}\cot\f{\pi\ell}{nL} \Big( (6h+5)\cos^2\f{\pi\ell}{L}-2\Big)                                     \\
&& \phantom{S_{n,\mX}^\oloop=}
      +2n^2 \f{\sin^2\f{\pi\ell}{L}}{\sin^2\f{\pi\ell}{nL}}\Big( h(54h+19)\cos^2\f{\pi\ell}{L}\cos^2\f{\pi\ell}{nL}
                                                                +(26h+9)\cos^2\f{\pi\ell}{L}
                                                                +4h\sin^2\f{\pi\ell}{nL} \Big)                                   \nn\\
&& \phantom{S_{n,\mX}^\oloop=}
     -12n \f{\sin^3\f{\pi\ell}{L}}{\sin^3\f{\pi\ell}{nL}}\cos\f{\pi\ell}{L}\cos\f{\pi\ell}{nL} \Big( h(6h+1)\cos^2\f{\pi\ell}{nL}+8h+3\Big) \nn\\
&& \phantom{S_{n,\mX}^\oloop=}
    +\f{\sin^4\f{\pi\ell}{L}}{\sin^4\f{\pi\ell}{nL}}\Big( h(18h+5)\cos^4\f{\pi\ell}{nL}+2(22h+9)\cos^2\f{\pi\ell}{nL}-4h+9\Big)
\bigg)-1\bigg]q^2 +O(q^3) \bigg\} + O(q^{2h}),  \nn
\eea
with $q=\ep^{-2\pi\b/L} \ll 1$. Taking $n \to 1$ limit, we get the one-loop holographic entanglement entropy
\be
S_{\mX}^\oloop = \Big( 1-\f{\pi\ell}{L}\cot\f{\pi\ell}{L} \Big) 2 q^h \big( h +(h+1)q +(h+2)q^2+O(q^3) \big) + O(q^{2h}).
\ee

Then we calculate the R\'enyi entropy in the CFT side. For the vacuum conformal family we have the holomorphic part of the density matrix
\be
\r_\vacu = \uum \vac +  \f{q^2}{\a_T}|T\rag\lag T|  + O(q^3).
\ee
Considering the contributions of the conformal family of a chiral primary operator $\mX$ we have the density matrix
\be
\r = \r_\vacu + \r_\mX,
\ee
with
\be
\r_\mX = q^h \Big(  \f{1}{\a_\mX}|\mX \rag \lag \mX |
                   + \f{q}{\a_{\p\mX}}|\p\mX \rag \lag \p\mX |
                   + \f{q^2}{\a_\mY}|\mY \rag \lag \mY |
                   + \f{q^2}{\a_{\p^2\mX}}|\p^2\mX \rag \lag \p^2\mX | + O(q^3)  \Big).
\ee
Then we get the R\'enyi entropy
\bea
&& S_{n,\mX}= -\f{n q^h}{n-1} \Big[ \Big( \f{\lag\mX(\inf)\mX(0)\rag_{\mC^n}}{\a_\mX} -1 \Big)
   +\Big( \f{\lag\p\mX(\inf)\p\mX(0)\rag_{\mC^n}}{\a_{\p\mX}} - 1 \Big)q
   +\Big( \f{\lag\p^2\mX(\inf)\p^2\mX(0)\rag_{\mC^n}}{\a_{\p^2\mX}} \nn\\
&& \phantom{S_{n,\mX}=}
          +\f{\lag\mY(\inf)\mY(0)\rag_{\mC^n}}{\a_{\mY}}
          -n\f{\lag T(\inf)T(0)\rag_{\mC^n}\lag\mX(\inf)\mX(0)\rag_{\mC^n}}{\a_{T}\a_{\mX}}
          +\sum_{j=1}^{n-1}\f{\lag T(\inf)T(0)\mX(\inf_j)\mX(0_j)\rag_{\mC^n}}{\a_{T}\a_{\mX}}
          -1 \Big)q^2 \nn\\
&& \phantom{S_{n,\mX}=}
   +O(q^3) \Big] + O(q^{2h}).
\eea
We need the correlation functions
\bea
&& \f{\lag\mX(\inf)\mX(0)\rag_{\mC^n}}{\a_\mX}=\f{1}{n^{2h}}\bigg(\f{\sin\f{\pi\ell}{L}}{\sin\f{\pi\ell}{nL}}\bigg)^{2h},  \nn\\
&& \f{\lag\p\mX(\inf)\p\mX(0)\rag_{\mC^n}}{\a_{\p\mX}}=
     \f{2}{n^{2h+2}} \bigg( \f{\sin\f{\pi\ell}{L}}{\sin\f{\pi\ell}{nL}}\bigg)^{2h}
     \bigg( n^2h\cos^2\f{\pi\ell}{L}
           -nh\sin\f{2\pi\ell}{L}\cot\f{\pi\ell}{nL}
           +\f{\sin^2\f{\pi\ell}{L}}{\sin^2\f{\pi\ell}{nL}} \Big(h\cos^2\f{\pi\ell}{nL}+\f12\Big)  \bigg),                                \nn\\
&& \f{\lag\p^2\mX(\inf)\p^2\mX(0)\rag_{\mC^n}}{\a_{\p^2\mX}}
      =\f{1}{2h+1}\f{1}{n^{2h+4}}\bigg( \f{\sin\f{\pi\ell}{L}}{\sin\f{\pi\ell}{nL}}\bigg)^{2h} \bigg[
          n^4h \Big( 2(h+1)\cos^2\f{\pi\ell}{L}-1\Big)^2                                                                                  \nn\\
&& \phantom{\f{\lag\p^2\mX(\inf)\p^2\mX(0)\rag_{\mC^n}}{\a_{\p^2\mX}}=}
         -2n^3h(2h+1) \sin\f{2\pi\ell}{L}\cot\f{\pi\ell}{nL} \Big( 2(h+1)\cos^2\f{\pi\ell}{L}-1\Big)                                      \nn\\
&& \phantom{\f{\lag\p^2\mX(\inf)\p^2\mX(0)\rag_{\mC^n}}{\a_{\p^2\mX}}=}
         +2n^2 \f{\sin^2\f{\pi\ell}{L}}{\sin^2\f{\pi\ell}{nL}}\Big( 2h\cos^2\f{\pi\ell}{nL}+1\Big)
                                                              \Big( (6h^2+6h+1)\cos^2\f{\pi\ell}{L}-h\Big)                                \nn\\
&& \phantom{\f{\lag\p^2\mX(\inf)\p^2\mX(0)\rag_{\mC^n}}{\a_{\p^2\mX}}=}
        -4n(2h+1)\f{\sin^3\f{\pi\ell}{L}}{\sin^3\f{\pi\ell}{nL}}\cos\f{\pi\ell}{L}\cos\f{\pi\ell}{nL}
                 \Big( 2h^2\cos^2\f{\pi\ell}{nL}+3h+1\Big)\\
&& \phantom{\f{\lag\p^2\mX(\inf)\p^2\mX(0)\rag_{\mC^n}}{\a_{\p^2\mX}}=}
        +\f{\sin^4\f{\pi\ell}{L}}{\sin^4\f{\pi\ell}{nL}}\Big( 4h^3\cos^4\f{\pi\ell}{nL}+2(6h^2+4h+1)\cos^2\f{\pi\ell}{nL}+3h+1\Big)\bigg], \nn\\
&& \f{\lag\mY(\inf)\mY(0)\rag_{\mC^n}}{\a_{\mY}} = \f{1}{n^{2h+4}}\bigg( \f{\sin\f{\pi\ell}{L}}{\sin\f{\pi\ell}{nL}}\bigg)^{2h}
        \bigg[ \f{(n^2-1)^2}{18(2h+1)}\sin^4\f{\pi\ell}{L}
               \Big( (2h+1)c+2h(8h-5) \Big) + \f{\sin^4\f{\pi\ell}{L}}{\sin^4\f{\pi\ell}{nL}} \bigg],                                      \nn\\
&& \f{\lag T(\inf)T(0)\mX(\inf_j)\mX(0_j)\rag_{\mC^n}}{\a_{T}\a_{\mX}}
      = \f{1}{n^{2h+4}}\bigg( \f{\sin\f{\pi\ell}{L}}{\sin\f{\pi\ell}{nL}}\bigg)^{2h} \sin^4\f{\pi\ell}{L}
        \bigg[ \f{(n^2-1)^2c}{18} + \f{1}{\sin^4\f{\pi\ell}{nL}}  \nn\\
&& \phantom{\f{ \lag T(\inf)T(0)\mX(\inf_j)\mX(0_j)\rag_{\mC^n}}{\a_{T}\a_{\mX}}=}
               +\f{(n^2-1)h}{3}\sin^2\f{\pi\ell}{nL}
                \f{\cos\f{2\pi j}{n}\cos\f{2\pi\ell}{nL}-1}{\sin^2\f{\pi j}{n}\lt(\cos\f{2\pi j}{n}-\cos\f{2\pi\ell}{nL}\rt)^2} \bigg] +O\Big(\f{1}{c}\Big). \nn
\eea
Taking the large $c$ limit we reproduce the one-loop gravity result (\ref{SnXgravityChiral}).

\section{Conclusion and discussion}\label{conanddis}

In this paper we have considered the contributions of a general non-vacuum conformal family to the R\'enyi mutual information of two intervals on complex plane and the R\'enyi entropy of one interval on torus in two-dimensional CFT. The primary operator of the conformal family can be either non-chiral or chiral. We got the results to the orders higher than those in literature, and found matches of gravity and CFT results.

We have only considered the contributions of one non-vacuum conformal family, and this is not complete for a concrete CFT. The algebra of the operators in the vacuum conformal family and one non-vacuum family is not close. For example at level $2\D$ there may be a new conformal family with primary operator
\be
\mO = (\mX\mX) + \cdots.
\ee
To make the result in this paper meaningful, we have to require that the scaling dimension $\D$ of the primary operator cannot be too small. For two intervals on complex plane we need $\D>1$, and for one interval on torus we need $\D>2$. For the contributions of a primary operators with a smaller scaling dimension and the contributions of more than two non-vacuum conformal families, further investigations are needed.

\section*{Acknowledgment}

The author would like to thank Bin Chen for valuable discussions and thank Peking University for hospitality.
The author thanks Matthew Headrick for his Mathematica code \emph{Virasoro.nb} that could be downloaded at \url{http://people.brandeis.edu/~headrick/Mathematica/index.html}. The work is in part supported by the ERC Starting Grant 637844-HBQFTNCER.

\appendix

\section{Review of non-vacuum conformal family}\label{rev}

In this appendix we review some properties of the non-vacuum conformal family that are useful for this paper, including the conformal family of a non-chiral primary operator and the conformal family of a chiral primary operator. Details can be found in \cite{DiFrancesco:1997nk,Blumenhagen:2009zz}, or can be easily derived from the results therein.

\subsection{Non-chiral primary operator}

The one-loop partition function of the vacuum conformal family is
\be \label{Zvac}
Z_\vacu = \prod_{k=2}^{+\inf} \f{1}{(1-q^k)(1-\bar q^k)}.
\ee
Considering the contributions of the conformal family of a non-chiral primary operator $\mX$, one has to multiply the result (\ref{Zvac}) by
\be
Z_\mX = 1 + \f{q^h\bar q^{\bar h}}{(1-q)(1-\bar q)}.
\ee
The non-chiral primary operator $\mX$ has conformal weights $(h,\bar h)$ with $h \neq 0$ and $\bar h \neq 0$. One has the scaling dimension $\D=h+\bar h$ and the spin $s=h-\bar h$. As usual, we require that $s$ is an integer or a half integer. In the conformal family of $\mX$, the operators can be written as quasiprimary operators and their derivatives. At level $(h+2,\bar h)$ and level $(h,\bar h+2)$ there are quasiprimary operators, respectively,
\be
\mY=(T\mX)-\f{3}{2(2h+1)}\p^2\mX, ~~
\mZ=(\bar T \mX)-\f{3}{2(2\bar h+1)}\bar \p^2\mX,
\ee
with the normalization factors being
\be
\a_\mY=\f{(2h+1)c+2h(8h-5)}{2(2h+1)}\a_\mX, ~~
\a_\mZ=\f{(2\bar h+1)c+2\bar h(8\bar h-5)}{2(2\bar h+1)}\a_\mX.
\ee
Note that $\a_\mX$ is the normalization factor of $\mX$, and that we consider a CFT with equaling holomorphic and anti-holomorphic central charges $c=\bar c$. Under a general conformal transformation $z\to f(z)$, $\bar z\to \bar f(\bar z)$, the primary operator $\mX$ transforms as
\be
\mX(z,\bar z) = f'^h\bar f'^{\bar h} \mX(f,\bar f),
\ee
and the quasiprimary operators $\mY$, $\mZ$ transform as
\bea
&& \mY(z,\bar z) = f'^{h+2}\bar f'^{\bar h}\mY(f,\bar f) + \f{(2h+1)c+2h(8h-5)}{12(2h+1)}s f'^h\bar f'^{\bar h} \mX(f,\bar f), \nn\\
&& \mZ(z,\bar z) = f'^{h}\bar f'^{\bar h+2}\mZ(f,\bar f)
                 + \f{(2\bar h+1)c+2\bar h(8\bar h-5)}{12(2\bar h+1)}f'^h\bar s \bar f'^{\bar h} \mX(f,\bar f),
\eea
with the Schwarzian derivatives
\be
s(z)=\f{f'''(z)}{f'(z)}-\f32 \bigg( \f{f''(z)}{f'(z)} \bigg)^2, ~~
\bar s(\bar z)=\f{\bar f'''(\bar z)}{\bar f'(\bar z)}-\f32 \bigg( \f{\bar f''(\bar z)}{\bar f'(\bar z)} \bigg)^2.
\ee

\subsection{Chiral primary operator}

For the conformal family of a chiral primary operator $\mX$ that has conformal weights $(h,0)$ with $h \neq 0$, one has to multiply the result (\ref{Zvac}) by
\be
Z_\mX = 1 + \f{q^h}{1-q}.
\ee
The scaling dimension is $\D=h$, and the spin $s=h$ is an integer or a half integer. At level $(h+2,0)$ there is quasiprimary operator
\be
\mY=(T\mX)-\f{3}{2(2h+1)}\p^2\mX, ~~
\a_\mY=\f{(2h+1)c+2h(8h-5)}{2(2h+1)}\a_\mX.
\ee
Under a general conformal transformation $z\to f(z)$ the primary operator $\mX$ and quasiprimary operator $\mY$ transform as
\be
\mX(z) = f'^h\mX(f), ~~
\mY(z) = f'^{h+2}\mY(f) + \f{(2h+1)c+2h(8h-5)}{12(2h+1)}s f'^h\mX(f).
\ee

\providecommand{\href}[2]{#2}\begingroup\raggedright\endgroup


\begin{thebibliography}{10}

\bibitem{Calabrese:2004eu}
P.~Calabrese and J.~L. Cardy, ``{Entanglement entropy and quantum field
  theory},'' \href{http://dx.doi.org/10.1088/1742-5468/2004/06/P06002}{{\em J.
  Stat. Mech.} {\bfseries 0406} (2004) P06002},
\href{http://arxiv.org/abs/hep-th/0405152}{{\ttfamily arXiv:hep-th/0405152
  [hep-th]}}.

\bibitem{Ryu:2006bv}
S.~Ryu and T.~Takayanagi, ``{Holographic derivation of entanglement entropy
  from AdS/CFT},'' \href{http://dx.doi.org/10.1103/PhysRevLett.96.181602}{{\em
  Phys. Rev. Lett.} {\bfseries 96} (2006) 181602},
\href{http://arxiv.org/abs/hep-th/0603001}{{\ttfamily arXiv:hep-th/0603001
  [hep-th]}}.

\bibitem{Ryu:2006ef}
S.~Ryu and T.~Takayanagi, ``{Aspects of Holographic Entanglement Entropy},''
  \href{http://dx.doi.org/10.1088/1126-6708/2006/08/045}{{\em JHEP} {\bfseries
  0608} (2006) 045},
\href{http://arxiv.org/abs/hep-th/0605073}{{\ttfamily arXiv:hep-th/0605073
  [hep-th]}}.

\bibitem{Callan:1994py}
C.~G. Callan~Jr. and F.~Wilczek, ``{On geometric entropy},''
  \href{http://dx.doi.org/10.1016/0370-2693(94)91007-3}{{\em Phys. Lett.}
  {\bfseries B333} (1994) 55--61},
\href{http://arxiv.org/abs/hep-th/9401072}{{\ttfamily arXiv:hep-th/9401072
  [hep-th]}}.

\bibitem{Holzhey:1994we}
C.~Holzhey, F.~Larsen, and F.~Wilczek, ``{Geometric and renormalized entropy in
  conformal field theory},''
  \href{http://dx.doi.org/10.1016/0550-3213(94)90402-2}{{\em Nucl. Phys.}
  {\bfseries B424} (1994) 443--467},
\href{http://arxiv.org/abs/hep-th/9403108}{{\ttfamily arXiv:hep-th/9403108
  [hep-th]}}.

\bibitem{Maldacena:1997re}
J.~M. Maldacena, ``{The Large N limit of superconformal field theories and
  supergravity},'' \href{http://dx.doi.org/10.1023/A:1026654312961}{{\em Int.
  J. Theor. Phys.} {\bfseries 38} (1999) 1113--1133},
  \href{http://arxiv.org/abs/hep-th/9711200}{{\ttfamily arXiv:hep-th/9711200
  [hep-th]}}.
[Adv. Theor. Math. Phys. {\bf 2}, (1998) 231].

\bibitem{Gubser:1998bc}
S.~Gubser, I.~R. Klebanov, and A.~M. Polyakov, ``{Gauge theory correlators from
  noncritical string theory},''
  \href{http://dx.doi.org/10.1016/S0370-2693(98)00377-3}{{\em Phys. Lett.}
  {\bfseries B428} (1998) 105--114},
\href{http://arxiv.org/abs/hep-th/9802109}{{\ttfamily arXiv:hep-th/9802109
  [hep-th]}}.

\bibitem{Witten:1998qj}
E.~Witten, ``{Anti-de Sitter space and holography},'' {\em Adv. Theor. Math.
  Phys.} {\bfseries 2} (1998) 253--291,
\href{http://arxiv.org/abs/hep-th/9802150}{{\ttfamily arXiv:hep-th/9802150
  [hep-th]}}.

\bibitem{Aharony:1999ti}
O.~Aharony, S.~S. Gubser, J.~M. Maldacena, H.~Ooguri, and Y.~Oz, ``{Large N
  field theories, string theory and gravity},''
  \href{http://dx.doi.org/10.1016/S0370-1573(99)00083-6}{{\em Phys.Rept.}
  {\bfseries 323} (2000) 183--386},
\href{http://arxiv.org/abs/hep-th/9905111}{{\ttfamily arXiv:hep-th/9905111
  [hep-th]}}.

\bibitem{Fujita:2009kw}
M.~Fujita, W.~Li, S.~Ryu, and T.~Takayanagi, ``{Fractional Quantum Hall Effect
  via Holography: Chern-Simons, Edge States, and Hierarchy},''
  \href{http://dx.doi.org/10.1088/1126-6708/2009/06/066}{{\em JHEP} {\bfseries
  06} (2009) 066},
\href{http://arxiv.org/abs/0901.0924}{{\ttfamily arXiv:0901.0924 [hep-th]}}.

\bibitem{Headrick:2010zt}
M.~Headrick, ``{Entanglement R\'enyi entropies in holographic theories},''
  \href{http://dx.doi.org/10.1103/PhysRevD.82.126010}{{\em Phys. Rev.}
  {\bfseries D82} (2010) 126010},
\href{http://arxiv.org/abs/1006.0047}{{\ttfamily arXiv:1006.0047 [hep-th]}}.

\bibitem{Barrella:2013wja}
T.~Barrella, X.~Dong, S.~A. Hartnoll, and V.~L. Martin, ``{Holographic
  entanglement beyond classical gravity},''
  \href{http://dx.doi.org/10.1007/JHEP09(2013)109}{{\em JHEP} {\bfseries 1309}
  (2013) 109},
\href{http://arxiv.org/abs/1306.4682}{{\ttfamily arXiv:1306.4682 [hep-th]}}.

\bibitem{Faulkner:2013ana}
T.~Faulkner, A.~Lewkowycz, and J.~Maldacena, ``{Quantum corrections to
  holographic entanglement entropy},''
  \href{http://dx.doi.org/10.1007/JHEP11(2013)074}{{\em JHEP} {\bfseries 1311}
  (2013) 074},
\href{http://arxiv.org/abs/1307.2892}{{\ttfamily arXiv:1307.2892}}.

\bibitem{Brown:1986nw}
J.~D. Brown and M.~Henneaux, ``{Central charges in the canonical realization of
  asymptotic symmetries: an example from three-dimensional gravity},''
\href{http://dx.doi.org/10.1007/BF01211590}{{\em Commun. Math. Phys.}
  {\bfseries 104} (1986) 207--226}.

\bibitem{Maloney:2007ud}
A.~Maloney and E.~Witten, ``{Quantum Gravity Partition Functions in Three
  Dimensions},'' \href{http://dx.doi.org/10.1007/JHEP02(2010)029}{{\em JHEP}
  {\bfseries 1002} (2010) 029},
\href{http://arxiv.org/abs/0712.0155}{{\ttfamily arXiv:0712.0155 [hep-th]}}.

\bibitem{Yin:2007gv}
X.~Yin, ``{Partition Functions of Three-Dimensional Pure Gravity},''
  \href{http://dx.doi.org/10.4310/CNTP.2008.v2.n2.a1}{{\em Commun. Num. Theor.
  Phys.} {\bfseries 2} (2008) 285--324},
\href{http://arxiv.org/abs/0710.2129}{{\ttfamily arXiv:0710.2129 [hep-th]}}.

\bibitem{Giombi:2008vd}
S.~Giombi, A.~Maloney, and X.~Yin, ``{One-loop Partition Functions of 3D
  Gravity},'' \href{http://dx.doi.org/10.1088/1126-6708/2008/08/007}{{\em JHEP}
  {\bfseries 0808} (2008) 007},
\href{http://arxiv.org/abs/0804.1773}{{\ttfamily arXiv:0804.1773 [hep-th]}}.

\bibitem{Faulkner:2013yia}
T.~Faulkner, ``{The Entanglement R\'enyi Entropies of Disjoint Intervals in
  AdS/CFT},''
\href{http://arxiv.org/abs/1303.7221}{{\ttfamily arXiv:1303.7221 [hep-th]}}.

\bibitem{Chen:2014unl}
B.~Chen and J.-q. Wu, ``{Single interval R\'enyi entropy at low temperature},''
  \href{http://dx.doi.org/10.1007/JHEP08(2014)032}{{\em JHEP} {\bfseries 08}
  (2014) 032},
\href{http://arxiv.org/abs/1405.6254}{{\ttfamily arXiv:1405.6254 [hep-th]}}.

\bibitem{Calabrese:2010he}
P.~Calabrese, J.~Cardy, and E.~Tonni, ``{Entanglement entropy of two disjoint
  intervals in conformal field theory II},''
  \href{http://dx.doi.org/10.1088/1742-5468/2011/01/P01021}{{\em J. Stat.
  Mech.} {\bfseries 1101} (2011) P01021},
\href{http://arxiv.org/abs/1011.5482}{{\ttfamily arXiv:1011.5482 [hep-th]}}.

\bibitem{Hartman:2013mia}
T.~Hartman, ``{Entanglement Entropy at Large Central Charge},''
\href{http://arxiv.org/abs/1303.6955}{{\ttfamily arXiv:1303.6955 [hep-th]}}.

\bibitem{Chen:2013kpa}
B.~Chen and J.-j. Zhang, ``{On short interval expansion of R\'enyi entropy},''
  \href{http://dx.doi.org/10.1007/JHEP11(2013)164}{{\em JHEP} {\bfseries 1311}
  (2013) 164},
\href{http://arxiv.org/abs/1309.5453}{{\ttfamily arXiv:1309.5453 [hep-th]}}.

\bibitem{Cardy:2014jwa}
J.~Cardy and C.~P. Herzog, ``{Universal Thermal Corrections to Single Interval
  Entanglement Entropy for Two Dimensional Conformal Field Theories},''
  \href{http://dx.doi.org/10.1103/PhysRevLett.112.171603}{{\em Phys. Rev.
  Lett.} {\bfseries 112} (2014) 171603},
\href{http://arxiv.org/abs/1403.0578}{{\ttfamily arXiv:1403.0578 [hep-th]}}.

\bibitem{Chen:2013dxa}
B.~Chen, J.~Long, and J.-j. Zhang, ``{Holographic R\'enyi entropy for CFT with
  $W$ symmetry},'' \href{http://dx.doi.org/10.1007/JHEP04(2014)041}{{\em JHEP}
  {\bfseries 1404} (2014) 041},
\href{http://arxiv.org/abs/1312.5510}{{\ttfamily arXiv:1312.5510 [hep-th]}}.

\bibitem{Perlmutter:2013paa}
E.~Perlmutter, ``{Comments on R\'enyi entropy in AdS$_3$/CFT$_2$},''
  \href{http://dx.doi.org/10.1007/JHEP05(2014)052}{{\em JHEP} {\bfseries 05}
  (2014) 052},
\href{http://arxiv.org/abs/1312.5740}{{\ttfamily arXiv:1312.5740 [hep-th]}}.

\bibitem{Chen:2014kja}
B.~Chen, F.-y. Song, and J.-j. Zhang, ``{Holographic R\'enyi entropy in
  AdS$_3$/LCFT$_2$ correspondence},''
  \href{http://dx.doi.org/10.1007/JHEP03(2014)137}{{\em JHEP} {\bfseries 1403}
  (2014) 137},
\href{http://arxiv.org/abs/1401.0261}{{\ttfamily arXiv:1401.0261 [hep-th]}}.

\bibitem{Beccaria:2014lqa}
M.~Beccaria and G.~Macorini, ``{On the next-to-leading holographic entanglement
  entropy in $AdS_{3}/CFT_{2}$},''
  \href{http://dx.doi.org/10.1007/JHEP04(2014)045}{{\em JHEP} {\bfseries 1404}
  (2014) 045},
\href{http://arxiv.org/abs/1402.0659}{{\ttfamily arXiv:1402.0659 [hep-th]}}.

\bibitem{Long:2014oxa}
J.~Long, ``{Higher Spin Entanglement Entropy},''
  \href{http://dx.doi.org/10.1007/JHEP12(2014)055}{{\em JHEP} {\bfseries 12}
  (2014) 055},
\href{http://arxiv.org/abs/1408.1298}{{\ttfamily arXiv:1408.1298 [hep-th]}}.

\bibitem{Chen:2014ehg}
B.~Chen and J.-q. Wu, ``{Universal relation between thermal entropy and
  entanglement entropy in conformal field theories},''
  \href{http://dx.doi.org/10.1103/PhysRevD.91.086012}{{\em Phys. Rev.}
  {\bfseries D91} (2015) 086012},
\href{http://arxiv.org/abs/1412.0761}{{\ttfamily arXiv:1412.0761 [hep-th]}}.

\bibitem{Chen:2014hta}
B.~Chen and J.-q. Wu, ``{Large interval limit of R\'enyi entropy at high
  temperature},'' \href{http://dx.doi.org/10.1103/PhysRevD.92.126002}{{\em
  Phys. Rev.} {\bfseries D92} (2015) 126002},
\href{http://arxiv.org/abs/1412.0763}{{\ttfamily arXiv:1412.0763 [hep-th]}}.

\bibitem{Chen:2015kua}
B.~Chen and J.-q. Wu, ``{Holographic calculation for large interval R\'enyi
  entropy at high temperature},''
  \href{http://dx.doi.org/10.1103/PhysRevD.92.106001}{{\em Phys. Rev.}
  {\bfseries D92} (2015) 106001},
\href{http://arxiv.org/abs/1506.03206}{{\ttfamily arXiv:1506.03206 [hep-th]}}.

\bibitem{Chen:2015uia}
B.~Chen, J.-q. Wu, and Z.-c. Zheng, ``{Holographic R\'enyi entropy of single
  interval on torus: with W symmetry},''
  \href{http://dx.doi.org/10.1103/PhysRevD.92.066002}{{\em Phys. Rev.}
  {\bfseries D92} (2015) 066002},
\href{http://arxiv.org/abs/1507.00183}{{\ttfamily arXiv:1507.00183 [hep-th]}}.

\bibitem{Chen:2015uga}
B.~Chen and J.-q. Wu, ``{1-loop partition function in AdS$_{3}$/CFT$_{2}$},''
  \href{http://dx.doi.org/10.1007/JHEP12(2015)109}{{\em JHEP} {\bfseries 12}
  (2015) 109},
\href{http://arxiv.org/abs/1509.02062}{{\ttfamily arXiv:1509.02062 [hep-th]}}.

\bibitem{Zhang:2015hoa}
J.-j. Zhang, ``{Holographic R\'enyi entropy for two-dimensional
  $\mathcal{N}=(1,1)$ superconformal field theory},''
  \href{http://dx.doi.org/10.1007/JHEP12(2015)027}{{\em JHEP} {\bfseries 1512}
  (2015) 027},
\href{http://arxiv.org/abs/1510.01423}{{\ttfamily arXiv:1510.01423 [hep-th]}}.

\bibitem{Li:2016pwu}
Z.~Li and J.-j. Zhang, ``{On one-loop entanglement entropy of two short
  intervals from OPE of twist operators},''
  \href{http://dx.doi.org/10.1007/JHEP05(2016)130}{{\em JHEP} {\bfseries 1605}
  (2016) 130},
\href{http://arxiv.org/abs/1604.02779}{{\ttfamily arXiv:1604.02779 [hep-th]}}.

\bibitem{Chen:2016uvu}
B.~Chen and J.-q. Wu, ``{Higher spin entanglement entropy at finite temperature
  with chemical potential},''
  \href{http://dx.doi.org/10.1007/JHEP07(2016)049}{{\em JHEP} {\bfseries 07}
  (2016) 049},
\href{http://arxiv.org/abs/1604.03644}{{\ttfamily arXiv:1604.03644 [hep-th]}}.

\bibitem{Chen:2016lbu}
B.~Chen, J.-B. Wu, and J.-j. Zhang, ``{Short interval expansion of R\'enyi
  entropy on torus},'' \href{http://dx.doi.org/10.1007/JHEP08(2016)130}{{\em
  JHEP} {\bfseries 08} (2016) 130},
\href{http://arxiv.org/abs/1606.05444}{{\ttfamily arXiv:1606.05444 [hep-th]}}.

\bibitem{Li:2016qbo}
Z.~Li and J.-j. Zhang, ``{Holographic R\'enyi entropy for two-dimensional
  $\mathcal{N}$=(2,2) superconformal field theory},''
\href{http://arxiv.org/abs/1611.00546}{{\ttfamily arXiv:1611.00546 [hep-th]}}.

\bibitem{DiFrancesco:1997nk}
P.~Di~Francesco, P.~Mathieu, and D.~Senechal, {\em {Conformal Field Theory}}.
\newblock Springer, New York, USA,
1997.
\newblock

\bibitem{Blumenhagen:2009zz}
R.~Blumenhagen and E.~Plauschinn, ``{Introduction to conformal field theory},''
\href{http://dx.doi.org/10.1007/978-3-642-00450-6}{{\em Lect. Notes Phys.}
  {\bfseries 779} (2009) 1--256}.

\end{thebibliography}

\end{document}